\documentclass{article}

\usepackage[left=2.5cm,
right=2.5cm,
top=3cm,
bottom=3cm,
headheight=40pt,
headsep=20pt,
a4paper]{geometry}

\usepackage{fancyhdr}       
\usepackage{XCharter}       
\usepackage[xcharter,bigdelims,vvarbb]{newtxmath} 
\usepackage[scaled=1.1]{zlmtt} 

\fancypagestyle{firstpage}{
  \fancyhf{} 
  \fancyhead[L]{\includegraphics[width=100pt]{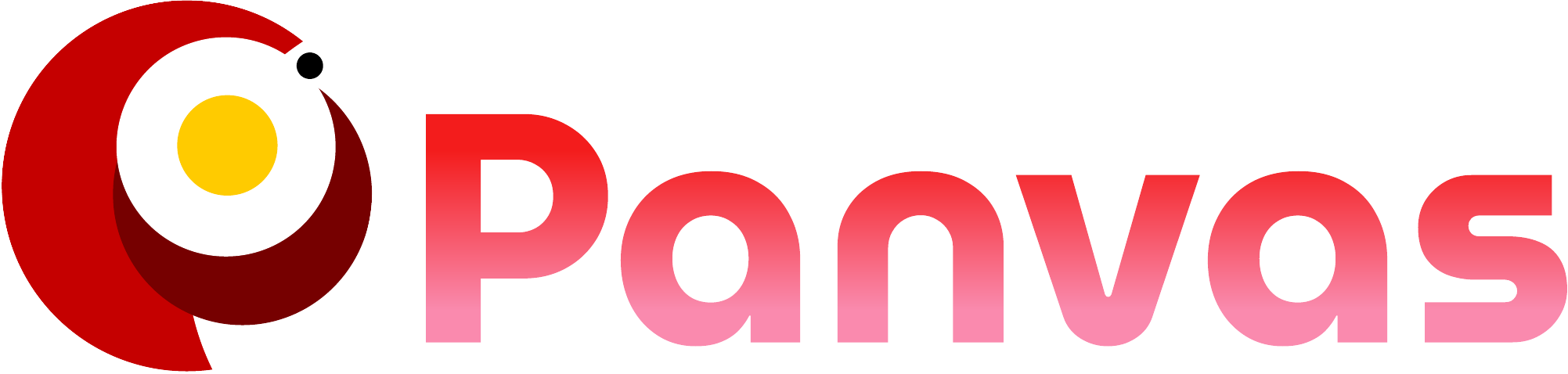}}
  \fancyhead[R]{\includegraphics[width=70pt]{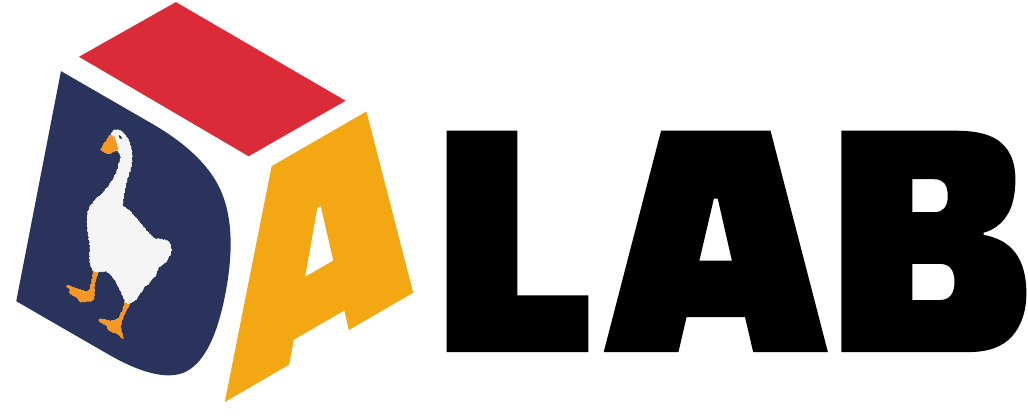}}
  \fancyfoot[C]{\thepage}

}

\newcommand{\eat}[1]{}

\usepackage{balance}
\usepackage{times}
\usepackage{latexsym}
\usepackage{amsfonts}
\usepackage{amsmath}
\usepackage{colortbl}
\usepackage{epsfig}
\usepackage{xspace}
\usepackage{graphicx}
\usepackage{comment}
\usepackage{booktabs}
\usepackage{multirow}
\usepackage{framed}
\usepackage{subcaption}

\usepackage{circledsteps}
\usepackage[utf8]{inputenc}
\usepackage{microtype}
\usepackage{inconsolata}
\usepackage[colorlinks=true, linkcolor=black, urlcolor=blue, citecolor=black]{hyperref}

\usepackage[ruled,linesnumbered]{algorithm2e}
\usepackage{algpseudocode}

\algnewcommand\algorithmicinput{\textbf{Input:}}
\algnewcommand\algorithmicoutput{\textbf{Output:}}
\algnewcommand\Input{\item[\algorithmicinput]}
\algnewcommand\Output{\item[\algorithmicoutput]}

\algdef{SE}[DOWHILE]{Do}{doWhile}{\algorithmicdo}[1]{\algorithmicwhile\ #1}%

\definecolor{green}{RGB}{0,128,0}

\definecolor{yellow}{RGB}{255,200,18}

\newcommand{\stab}{\vspace{1.2ex}\noindent}

\newcommand{\bi}{\begin{itemize}}
\newcommand{\ei}{\end{itemize}}

\newcommand{\be}{\begin{enumerate}}
\newcommand{\ee}{\end{enumerate}}
\newcommand{\beqn}{\begin{eqnarray*}}
\newcommand{\eeqn}{\end{eqnarray*}}

\newcommand{\stitle}[1]{\stab\noindent{\bf #1}}
\newcommand{\etitle}[1]{\vspace{1mm}\noindent{\underline{\em #1}}}

\newcommand{\aka}{\emph{a.k.a.}\xspace}

\usepackage{tabu}                      
\usepackage{booktabs}                  
\usepackage{lipsum}                    
\usepackage{mwe}                       

\usepackage{listings}

\usepackage{pifont}
\usepackage{tikz}
\usepackage{enumitem}

\newcommand{\sys}{\texttt{\textbf{PANVAS}}\xspace}

\title{Rise of the Community Champions: \\ From Reviewer Crunch to Community Power}

\author{%
  Changlun Li\thanks{
    Preprint.  Work in Progress. We welcome discussion and collaboration.
    } \\
  HKUST(GZ)\\
  cli942@connect.hkust-gz.edu.cn
  \and
  Yao Shi \\
  HKUST(GZ) \\
  yshi236@connect.hkust-gz.edu.cn
  \and
  Yuyu Luo \\
  HKUST(GZ) \\
  yuyuluo@hkust-gz.edu.cn
  \and
  Nan Tang \\
  HKUST(GZ) \\
  nantang@hkust-gz.edu.cn
}

\date{}

\begin{document}

\maketitle
\thispagestyle{firstpage}

\begin{figure}[h!]
	\centering
	\includegraphics[width=.9\textwidth]{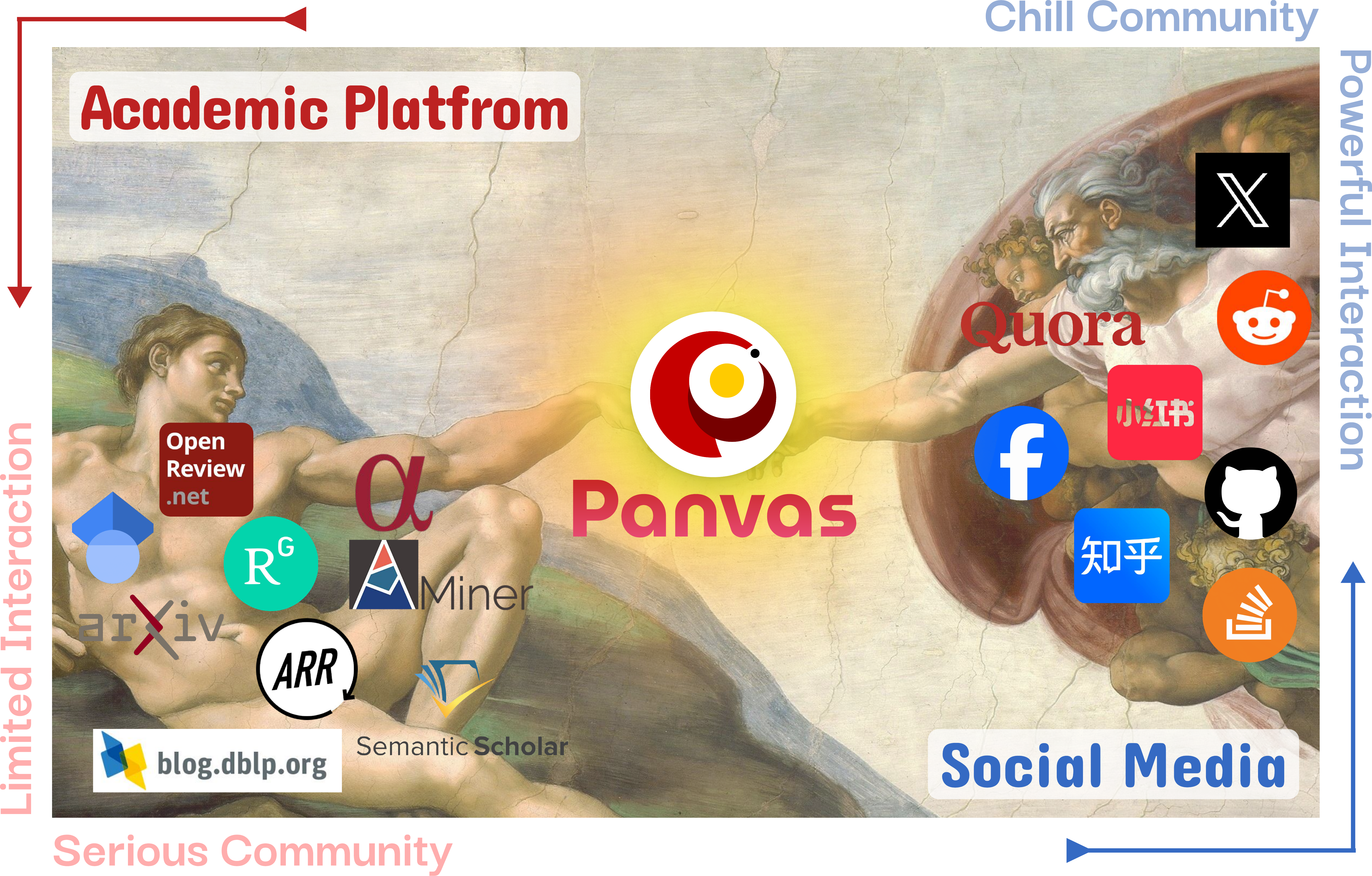}
\end{figure}

\begin{abstract}
Academic publishing is facing a crisis driven by exponential growth in submissions and an overwhelmed peer review system, leading to inconsistent decisions and a severe reviewer shortage. This paper introduces Panvas, a platform that reimagines academic publishing as a continuous, community-driven process. Panvas addresses these systemic failures with a novel combination of economic incentives (paid reviews) and rich interaction mechanisms (multi-dimensional ratings, threaded discussions, and expert-led reviews). By moving beyond the traditional accept/reject paradigm and integrating paper hosting with code/data repositories and social networking, Panvas fosters a meritocratic environment for scholarly communication and presents a radical rethinking of how we evaluate and disseminate scientific knowledge. We present the system design, development roadmap, and a user study plan to evaluate its effectiveness. 
\end{abstract}

\section{Background}

The academic publishing landscape has reached a breaking point, characterized by a dramatic surge in submissions and an evolving approach to dissemination. Platforms like arXiv~\cite{boldt2011extending} receive thousands of paper submissions daily.  Conference submission rates have exploded~\cite{papercopilot}—NeurIPS submissions have increased 2.5 times in just the last five years (2019-2024), a trend further accelerated by advancements in GPT and other AI technologies~\cite{achiam2023gpt,zhang2025aflow,spo,aot}. This overwhelming volume creates a system where truly innovative work struggles to gain visibility.
The peer review process—supposedly the gold standard for quality control—has become a critical bottleneck~\cite{Bucur2019PeerRR}. The sheer scale of research output demands new approaches to maintain quality and fairness in evaluation. The existing system, designed for a smaller, slower-paced academic world, is struggling to adapt.
The consequences of this broken system extend beyond administrative inefficiency. When brilliant ideas remain buried under submission volumes or face dismissive reviews, the entire scientific community loses potential breakthroughs. This systemic failure undermines the core purpose of academic publishing: to advance human knowledge through rigorous but fair evaluation of new ideas.
Current platforms fail to fully address these fundamental issues. Conference review systems remain time-constrained and episodic. Preprint servers like DBLP~\cite{dblp}, arXiv~\cite{arxiv}, ResearchGate~\cite{thelwall2017researchgate}, AMiner~\cite{Tang08KDD} offer limited community engagement. Review platforms like OpenReview~\cite{soergel2013open}, ACL Rolling Review (ARR)~\cite{dabre2022acl}, while more interactive, have sometimes devolved into hostile environments that discourage participation.

This growth coincides with a shift away from the traditional, strictly conference-centric model of academic publishing. The community is increasingly incorporating open review systems and integrating with social networks to facilitate broader discussion and faster dissemination of research.
For example, researchers are increasingly using social media platforms like X.com (\aka Twitter) and Reddit to share their work, and to solicit feedback from the community. AlphaXiv, a project supervised by Andrew Ng, Stanford AI Lab, integrates open discussion directly on arXiv papers~\cite{alphaxiv}. Leading AI institutions like Meta FAIR, OpenAI, and Google DeepMind now increasingly publish directly on their websites, bypassing the traditional peer review process entirely. 

In this paper, we introduce \sys—a portmanteau of \textit{Paper} and \textit{Canvas}—which offers a radical rethinking of academic publishing as an ongoing, community-driven conversation rather than a series of gatekeeping events as shown in the teaser figure.
We present evidence of the current system's failings, outline a vision for what academic publishing should become, and detail our solution to these entrenched problems in the following sections.

\section{Current Problems in Publication Review}

The academic review process is facing a multifaceted crisis. This section examines the key failures that undermine the effectiveness of current peer review systems, creating an environment where groundbreaking research struggles to gain recognition. As Yoshua Bengio, Chair of ICLR 2013, noted, the \textit{reviewer crunch} creates a severe bottleneck in allocating appropriate reviewers, particularly during peak conference reviewing periods. This scarcity of qualified reviewers leads to reduced review quality and undermines the perception of fairness in academic publishing. A decade later, these problems (P1-P3) have only intensified.

\stitle{P1: Inconsistent Decisions.}
The consequences of this crunch are starkly illustrated by the 2021 NeurIPS Consistency Experiment, as shown in Table~\ref{tab:inconsistency}. Program chairs duplicated 10\% of submissions, assigning them to two independent committees. The results were alarming: 203 out of 882, or 23\%, revisited papers received inconsistent decisions. Even more troubling, more than half of all papers recommended for spotlight presentations by either committee were rejected by the other (13/25 and 13/23). Even though there is no evidence that the decision process has become more or less noisy with increasing scale, this experiment formally confirmed what many researchers already suspected—the review process has become largely arbitrary, particularly for borderline papers.

\begin{table}[h!]
    \centering
    \caption{The decision inconsistency of NeurIPS 2021. The original classification counts in the column-wise, and the copied one counts in the row-wise. For example, there are 9 papers (underlined) that are recommended for Spotlight at original but classified as Poster at copy.}
    \label{tab:inconsistency}
    \begin{tabular}{|l|lllll|}
    \hline
    Original vs Copy & Oral & Spotlight & Poster & Reject & Withdrawn \\ \hline
    Oral             & 0    & 0         & 4      & 0      & 0         \\
    Spotlight        & 0    & 3         & \underline{9}      & 13     & 0         \\
    Poster           & 2    & 7         & 74     & 94     & 0         \\
    Reject           & 0    & 13        & 83     & 462    & 0         \\
    Withdrawn        & 0    & 0         & 0      & 0      & 118       \\ \hline
    \end{tabular}
\end{table}

\begin{figure}[t!]
    \centering
    \includegraphics[width=.9\textwidth]{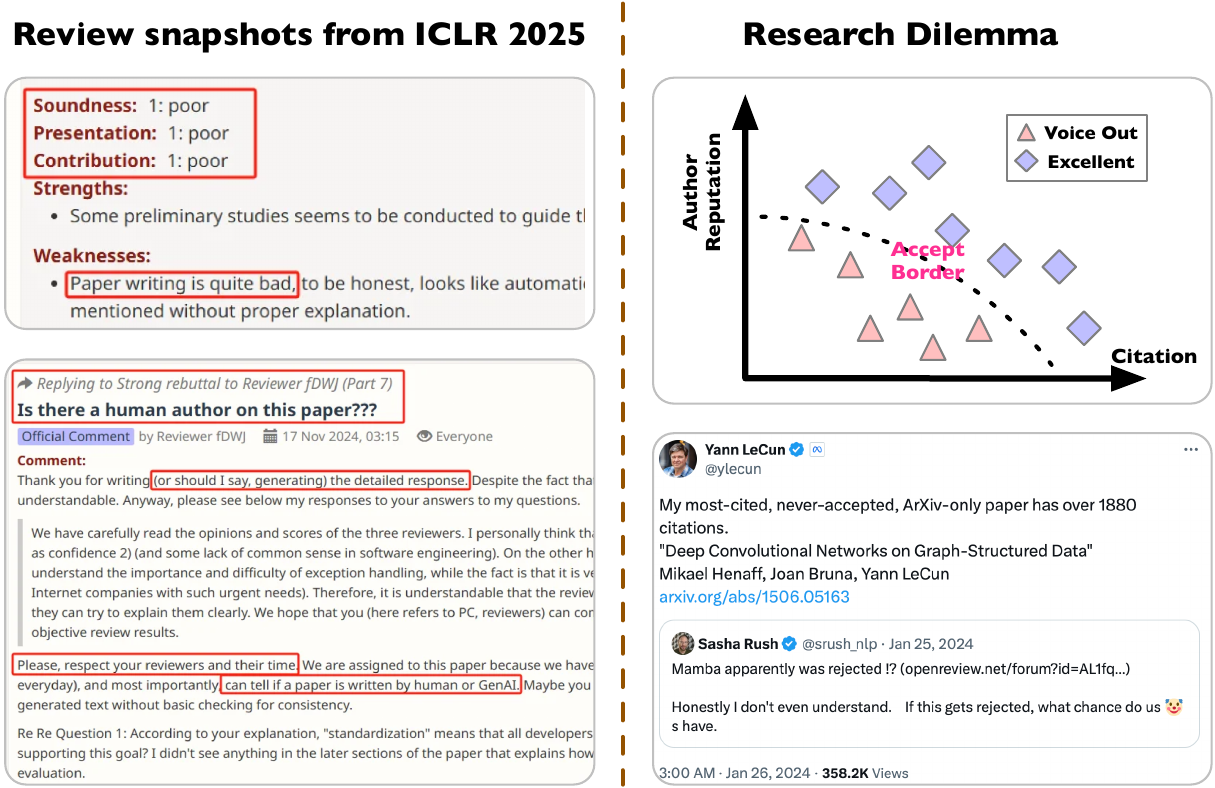}
    \caption{Real-world examples in current publication systems.}
    \label{fig:failures}
\end{figure}

\stitle{P2: Improfessional Reviews.}
While open review systems have been implemented for over a decade, significant challenges persist. Platforms like OpenReview and ACL Rolling Review, despite their aim to increase transparency and interaction, have faced criticism for fostering hostile environments and exhibiting instances of offensive commentary (see left panel of Figure~\ref{fig:failures}). While, some reviewers show their ignorance of the author's response during the rebuttal phase.
The CVPR 2025 review process also spotted that fully AI-assisted reviews were made by a few highly irresponsible reviewers. One of the co-authors of LoRA paper~\cite{hu2022lora}, Allen-Zhu Zeyuan, also criticized a similar issue in the current review process~\cite{AX2025-doge}. These observations highlight the need for a more pleasant and constructive review environment.

\stitle{P3: The Visibility Gap for Unconventional Research.}
Another critical problem lies in the inherent biases of the current system against unconventional or groundbreaking research that doesn't fit neatly into established paradigms. As illustrated in the right panel of Figure~\ref{fig:failures}, the Turing-awardee Yann LeCun stated that his most-cited, never-accepted paper is an arXiv-only paper.
There are more instances of research that, despite failing to gain acceptance at top conferences, ultimately achieve significant recognition and impact within the academic community.
Researchers pursuing such unconventional work face a significant ``visibility gap''. Traditional venues, with their emphasis on established norms and incremental advancements, may be ill-equipped to recognize the potential of truly disruptive ideas. This creates a dilemma for researchers: how to gain recognition and promote their work when the established gatekeepers fail to appreciate its value?  The current system lacks effective mechanisms to address this challenge, potentially stifling innovation and delaying the progress of scientific discovery.

The problems outlined above aren't merely inconveniences; they represent fundamental failures in how we evaluate and disseminate scientific knowledge. They demand not incremental changes but a complete reimagining of the academic publishing paradigm. 
\section{Critical Need and Our Vision}

The flaws in the current academic publishing system have a profound impact on research quality and fairness. The perception of credibility is damaged, potentially deterring researchers from contributing their best work. This section outlines the critical need to address these issues and presents a vision for a more effective and equitable publishing platform.

The current system's inconsistencies and inefficiencies can lead to the rejection of innovative work, hindering scientific progress. Researchers, particularly early-career scientists, may face career setbacks due to arbitrary review outcomes. The lack of constructive feedback and the potential for hostile interactions further discourage participation and create an uneven playing field.
Researchers envision a platform where they can submit their work at any time and receive trustworthy, constructive reviews. This ideal system would facilitate continuous open discussion, allowing papers to evolve based on community feedback. It would move beyond the binary accept/reject model, recognizing that scientific understanding develops over time. Furthermore, this visionary platform should combine the best elements of academic social networks like ResearchGate and AlphaXiv with the transparency of open review systems, while implementing safeguards against negativity. By fostering a culture of respectful criticism and merit-based visibility, we can create an environment where innovative research—even if unconventional—can receive the recognition it deserves regardless of author affiliation or reputation.

\begin{figure}[t!]
    \centering
    \includegraphics[width=.5\textwidth]{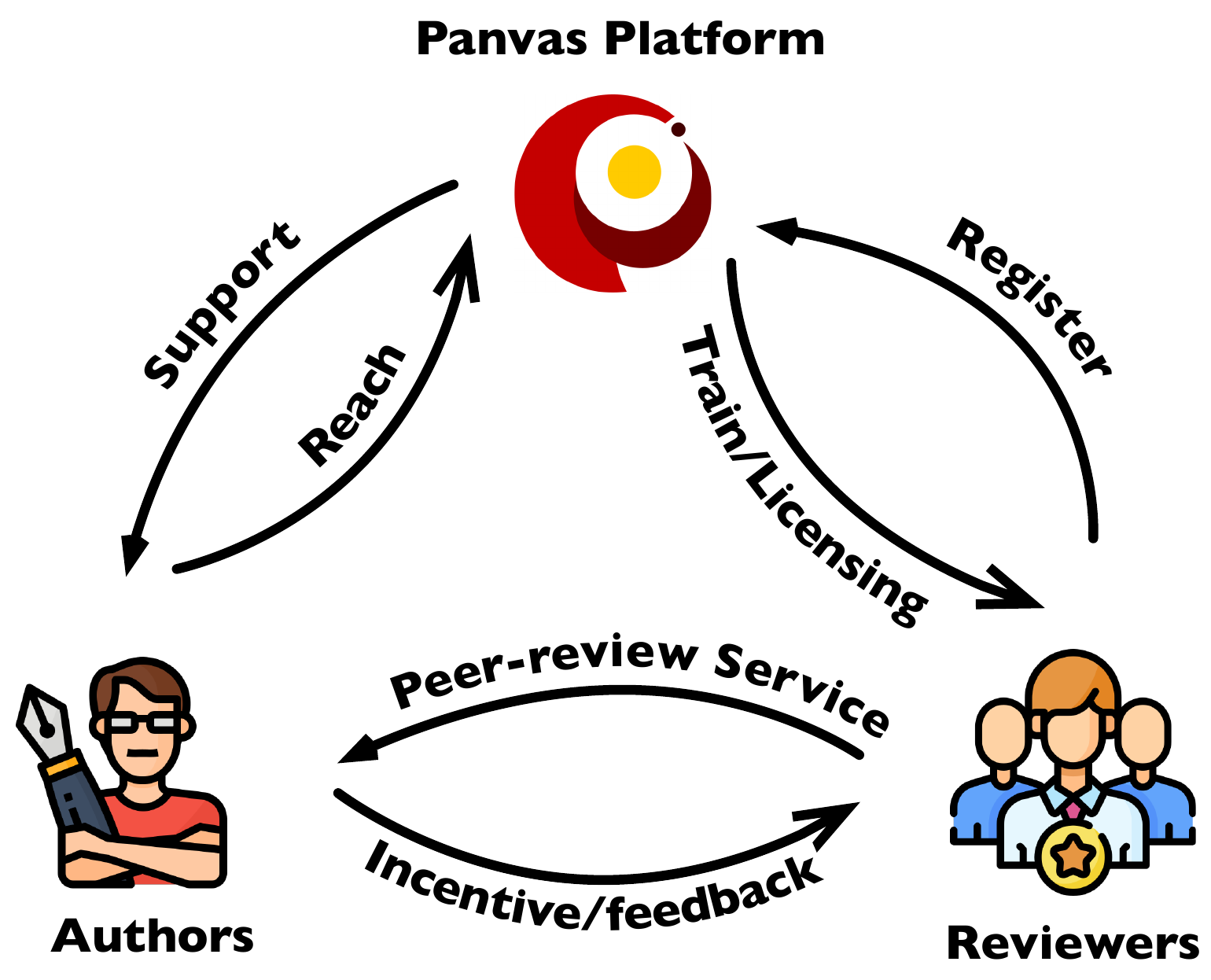}
    \caption{The ideal platform creates a structured ecosystem connecting authors and reviewers through multiple supportive pathways, addressing key failures in current peer review systems.}
    \label{fig:vision}
\end{figure}

\stitle{How Panvas Achieves the Vision?}
\sys addresses the challenges of academic publishing by creating a balanced ecosystem that benefits all stakeholders:

\etitle{Author Perspective.} 
Authors currently suffer from inconsistent decisions and limited visibility for unconventional work. \sys enables authors to submit work at any time, reach the platform through multiple channels, and receive ongoing support throughout the publication journey. Rather than facing binary accept/reject decisions that may stifle innovation, authors receive constructive feedback that helps evolve their work. This addresses the visibility gap for unconventional research by providing alternative pathways to recognition beyond traditional gatekeeping mechanisms.

\etitle{Reviewer Perspective.} 
The current review environment often creates arbitrary outcomes and can foster hostility. \sys transforms this experience by providing reviewers with proper training and licensing, ensuring they are equipped to provide trustworthy, constructive reviews. The bidirectional incentive/feedback mechanism acknowledges reviewers' contributions while maintaining quality standards. This structured approach helps overcome the inconsistent decisions and hostile environments that plague current systems.

\etitle{Platform Perspective.} 
The platform serves as the critical bridge in this ecosystem, embodying the desired venue characteristics. By combining elements from academic social networks with transparent review processes, \sys creates a comprehensive environment that facilitates continuous open discussion. The platform implements safeguards against negativity while fostering respectful criticism and merit-based visibility. As illustrated in Figure~\ref{fig:vision}, the platform connects all stakeholders through complementary pathways, creating a structured ecosystem that addresses the fundamental failures in current peer review systems.

\sys directly addresses the key problems in current systems: inconsistent decisions (\textbf{P1}) through better reviewer selection and training; hostile environments in open review (\textbf{P2}) through constructive feedback channels; and the visibility gap for unconventional research (\textbf{P3}) by creating multiple pathways for work to gain recognition beyond traditional gatekeeping mechanisms. As illustrated in Figure~\ref{fig:vision}, our vision centers on a platform that connects authors and reviewers through multiple complementary pathways. Authors can reach the platform to submit their work and receive support throughout the process. Reviewers register and receive appropriate training and licensing to ensure quality control. Between these stakeholders, a bidirectional relationship enables both peer-review service delivery and meaningful incentive/feedback mechanisms.
\section{Proposed Solution}

\sys offers a transformative approach to academic publishing, reimagining it as a paper-centric hub that integrates resources, data/models, and social networks. This creates a comprehensive ecosystem (see Figure~\ref{fig:view_b}) that addresses the limitations of existing systems.

\subsection{Reimagining Review and Engagement}

\sys fundamentally restructures the review process and fosters active community engagement.  We move beyond the simplistic accept/reject paradigm with several key innovations:

\stitle{Incentivized Review Process:}  Authors can offer rewards to attract qualified reviewers, and reviewers can bid for papers. This market-based system, inspired by recommendations from Oxford researchers, recognizes reviewing as valuable intellectual labor and incentivizes thorough, timely evaluations. Reviewer accreditation ensures transparency, while meta-reviews (community evaluation of reviews) maintain quality and constructive feedback.  Expertise matching ensures papers are reviewed by those genuinely qualified.

\begin{figure*}[t!]
	\centering
	\begin{subfigure}[t]{0.48\textwidth}
		\centering
		\includegraphics[width=\textwidth]{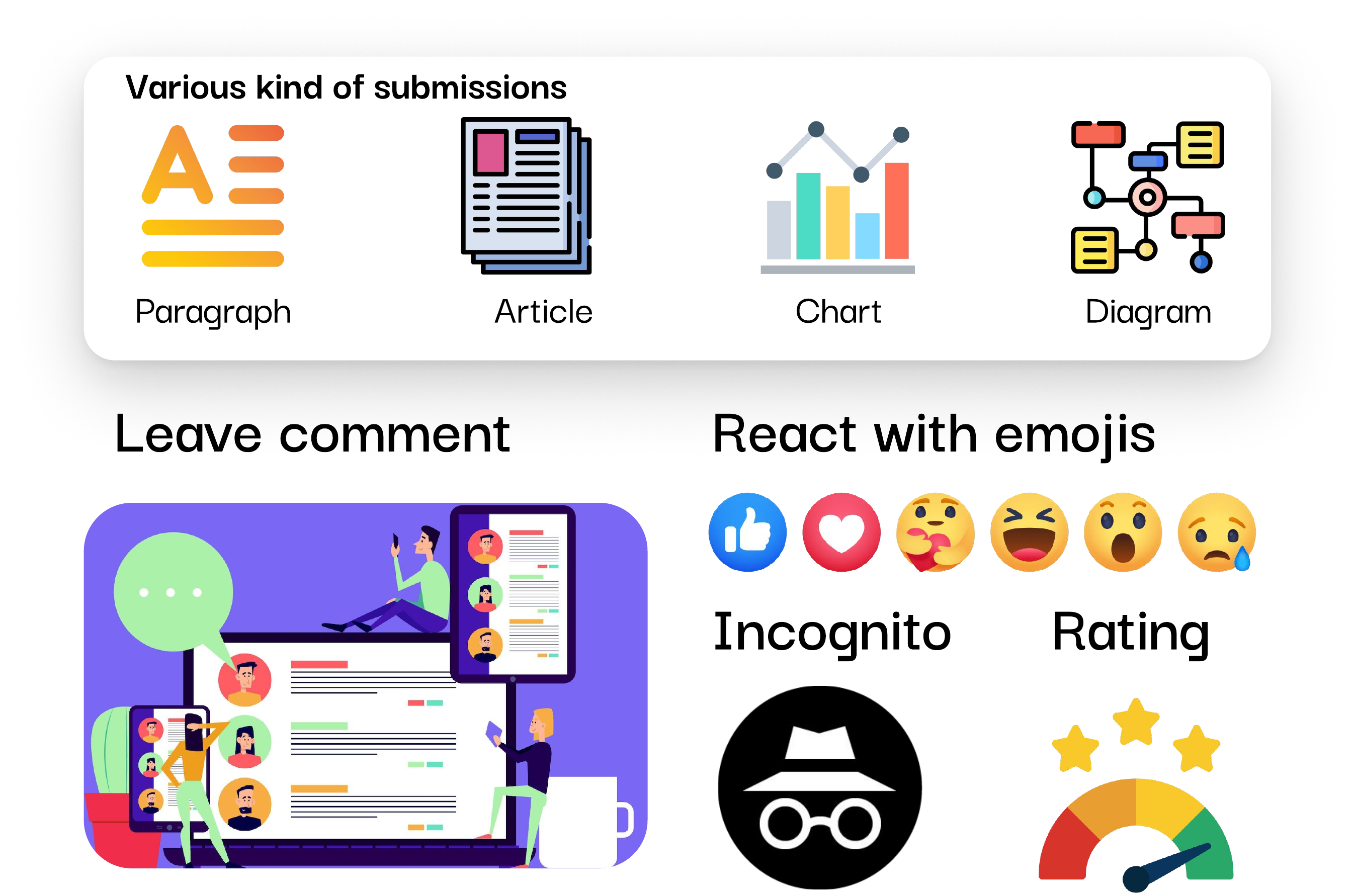}
		\caption{An overview of \sys's interaction design, showing how community engagement transforms the traditional review process into a continuous dialogue.}
		\label{fig:view_a}
	\end{subfigure}
	\hfill
	\begin{subfigure}[t]{0.48\textwidth}
		\centering
		\includegraphics[width=\textwidth]{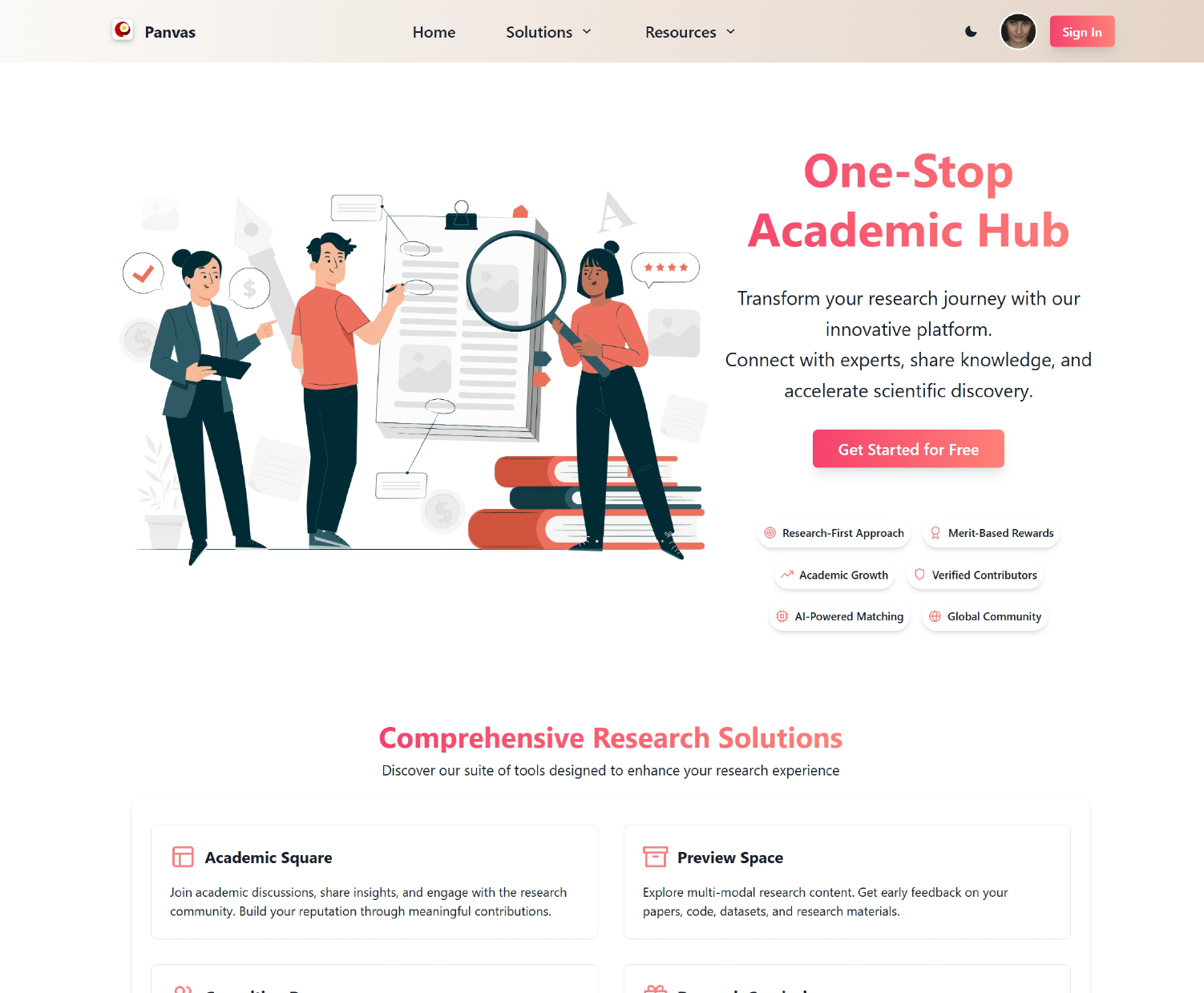}
		\caption{\sys currently envisions a front-end rendering version, which does not represent the final view.}
		\label{fig:view_b}
	\end{subfigure}
	\caption{Overview of \sys design and implementation}
	\label{fig:views}
\end{figure*}

\stitle{Rich Interaction Mechanisms:}  \sys transforms passive consumption into active engagement. As shown in Figure~\ref{fig:view_a}, Multi-dimensional rating systems replace binary judgments, allowing nuanced evaluation across criteria like originality, soundness, and impact. Threaded discussions attached to specific paper sections enable continuous dialogue, creating a ``living document'' that evolves with community feedback.  Reaction emojis provide lightweight feedback, capturing immediate impressions. Besides, a creative predictive voting feature allows users to ``bet'' on paper acceptance at target venues, creating a prediction market for research quality and leveraging collective intelligence to identify promising work early. 

\stitle{Fragmental submission:}  \sys allows authors to submit papers in a fragmental way, which is more flexible and convenient for authors. For example, an author can submit a paper in the form of a ``paragraph'' or a ``diagram'', and then gradually complete the paper. Charts can be uploaded and reviewed as separate panels, and then linked to the paper.

\subsection{Democratizing Academic Discourse and Ensuring Inclusivity}

\sys promotes equitable participation and addresses power imbalances that often characterize academic communities.  We recognize that traditional academic structures can create barriers to open and honest discourse, and we actively counter these through several key mechanisms:

\stitle{Incognito Mode:}  This feature allows users to provide feedback and participate in discussions without revealing their identity.  This is crucial for mitigating the fear of professional repercussions, particularly for junior researchers or those critiquing the work of established figures.  In traditional academic settings, power dynamics can discourage honest critique, as individuals may worry about negative consequences for their careers.  Incognito mode levels the playing field, allowing ideas to be evaluated on their merits, regardless of the source. It fosters a more open and critical environment where constructive criticism is encouraged, and diverse perspectives are valued.

\stitle{Merit-Based Evaluation:}  \sys is designed to prioritize the substance of contributions over the reputation or authority of the contributor.  This means that ideas are judged based on their quality, originality, and rigor, rather than on the author's credentials, affiliation, or prior publications.  This is a significant departure from traditional academic hierarchies, where established researchers often hold disproportionate influence.  By equalizing participation, \sys creates opportunities for students, early-career researchers, and individuals from underrepresented institutions to gain recognition for their work.  This fosters a more diverse and inclusive academic community where talent and innovation are recognized regardless of background.

\stitle{Community Moderation:}  While fostering open discussion, \sys also recognizes the need to maintain a civil and respectful environment.  Community moderation, supported by intelligent algorithms, helps to identify and address problematic behavior, such as personal attacks, harassment, or discriminatory language.  This is crucial for ensuring that all participants feel safe and welcome to contribute their ideas.  However, the moderation system is carefully designed to avoid stifling legitimate academic debate or suppressing dissenting opinions.  The goal is to strike a balance between promoting robust intellectual exchange and maintaining a constructive and inclusive atmosphere. This ensures that rigor and respect coexist, fostering a productive environment for scholarly communication.





\section{User Study Design}

To evaluate the effectiveness of \sys as a transformative academic publishing platform, we have designed a comprehensive user study centered around our novel collaboration workflow model. This section outlines our study design and methodology, with results to be incorporated in future work.

\subsection{The \sys Collaboration Workflow Model}

At the core of our user study is the \sys collaboration workflow—a tripartite ecosystem consisting of three primary roles: Freeman, Producer, and Consumer. This model represents a fundamental shift from traditional academic publishing relationships:

\begin{figure*}[t!]
	\centering
	\includegraphics[width=\textwidth]{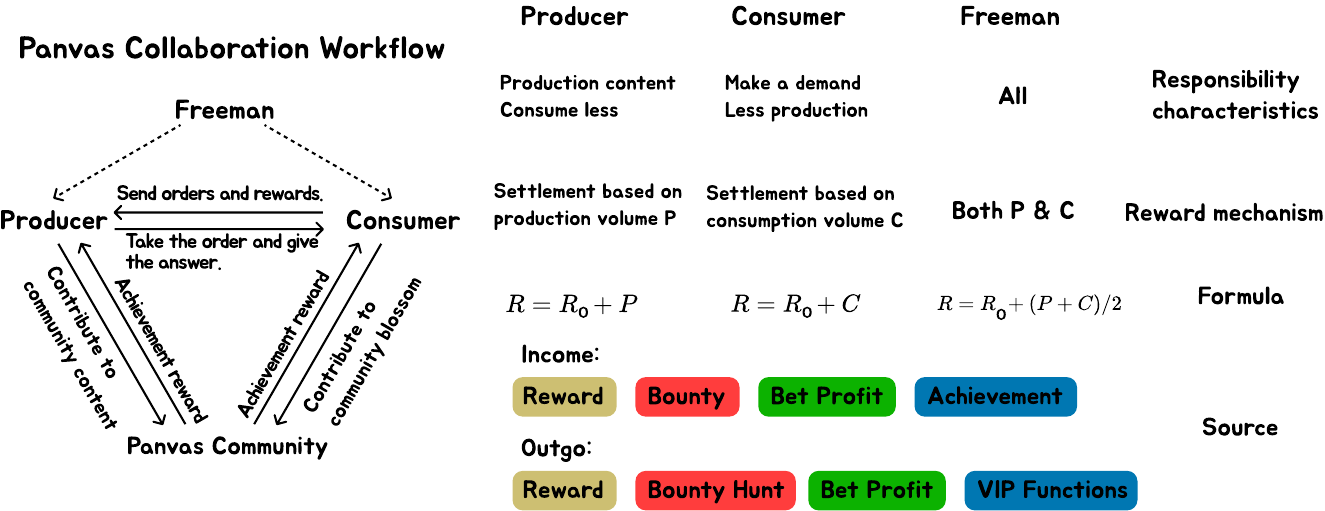}
 	\caption{The \sys collaboration workflow showing interactions between Freeman, Producer, Consumer, and the broader \sys community. This model underpins our incentive structure and reward mechanisms.}
	\label{fig:evalworkflow}
\end{figure*}

\textbf{Producer} focuses on creating content while consuming less. In academic terms, this corresponds primarily to authors who contribute papers and reviewers who produce detailed evaluations. Producers take orders and provide answers, with their settlement based on production volume (P), following the formula R = R$_0$ + P, where R$_0$ represents a base reward.

\textbf{Consumer} makes demands while producing less. These are readers, voters, and community members who engage with content through ratings, comments, and predictive voting. Their settlement is based on consumption volume (C), calculated as R = R$_0$ + C.

\textbf{Freeman} serves as a flexible and unrestricted role within the system. They can act as both producers and consumers, much like general users on a platform. They both send orders and distribute rewards, and have the ability to participate in content-related activities from multiple perspectives. This role is responsible for maintaining the overall balance of the system. Freemen utilize a reward mechanism that takes into account both production volume (P) and consumption volume (C), calculated by the formula R = R\(_0\)+ (P + C)/2. This balanced approach helps to ensure that the system functions smoothly and fairly, promoting a healthy interaction between different user groups.

The \sys Community sits at the intersection of these roles, where participants contribute to the community context and receive achievement rewards for their contributions. The community serves as both interaction space and value-creation mechanism, with a hybrid reward formula of R = R$_0$ + (P+C)/2 applied when participants embody both Producer and Consumer roles.

\subsection{Reward and Incentive Structure}

Our user study will evaluate the effectiveness of multiple income and outgo streams designed to incentivize quality participation. For income sources, the platform provides direct rewards as compensation for reviews and contributions, bounties for fulfilling specific community needs, bet profits from successful predictive voting on paper outcomes, and achievements in the form of recognition and rewards for milestone accomplishments. These income mechanisms are carefully balanced to reward different types of valuable participation without creating perverse incentives.

On the outgoing side, participants may invest in rewarding valuable contributions from others, engage in bounty hunting by spending resources to seek specific expertise or content, place stakes as bet profits on paper outcome predictions, and access premium VIP functions available to established participants. This economic ecosystem creates a self-sustaining cycle where value flows to quality contributions while maintaining system integrity through balanced incentives.

Future work will integrate detailed results and analysis, which are set to powerfully illustrate the effectiveness of \sys and the promising outcomes of our collaborative workflow model in transforming academic publishing.

\section{Conclusion}

In this paper, we introduced \sys as a revolutionary approach to address the deep-seated issues plaguing the academic publishing landscape.
Our vision for \sys centered around several key aspects. For authors, we envisioned a platform where they could submit their work at any time and receive constructive feedback, rather than facing the binary and potentially stifling accept/reject decisions of the traditional system. This would also help to bridge the visibility gap for unconventional research by providing alternative routes to recognition. Reviewers, on the other hand, would be provided with proper training and licensing, and a bidirectional incentive/feedback mechanism would be in place to ensure they are motivated to provide high  quality and trustworthy reviews. The platform itself would serve as a comprehensive ecosystem, integrating academic social networks and transparent review processes, while safeguarding against negativity.

However, several open questions remain. First, while our user study design provides a framework to evaluate the effectiveness of \sys, the actual results of the study are yet to be analyzed. It remains to be seen how well the proposed incentive structures will work in practice. Will the economic incentives lead to a significant improvement in review quality and reviewer engagement? Or will they introduce new forms of bias or gaming behavior? Second, the long term sustainability of the platform's community moderation system is a concern. As the user base grows, it may be challenging to maintain a civil and respectful environment while still allowing for robust academic debate. We need to further explore how to optimize the community moderation algorithms and ensure they are fair and effective. Third, the integration of \sys with existing academic infrastructure, such as other preprint servers and citation databases, requires careful consideration. 

To address these open questions, we expect the following solutions. Through the user study, we will collect and analyze data on user behavior, review quality, and community engagement. Based on these results, we can refine the incentive mechanisms, for example, by adjusting the reward weights for different types of contributions. Regarding community moderation, we plan to continuously improve the algorithms by incorporating more advanced natural language processing techniques to better detect and prevent inappropriate behavior. We also aim to establish clear moderation guidelines and train moderators to handle complex situations. For the integration with existing academic infrastructure, we will collaborate with other platforms and stakeholders to develop common standards and protocols, ensuring that \sys can co-exist and interact effectively with the broader academic ecosystem. Overall, we believe that with continuous improvement and innovation, \sys has the potential to revolutionize academic publishing and provide a more fair, efficient, and inclusive platform for scholarly communication. 

\newpage
\bibliographystyle{plainurl}
\bibliography{main}

\end{document}